\documentclass[a4paper,fleqn,usenatbib,usedcolumn]{mnras}

\usepackage[T1]{fontenc}
\usepackage{ae,aecompl}

\usepackage{graphicx}	% Including figure files
\usepackage{amsmath}	% Advanced maths commands
\usepackage{amssymb}	% Extra maths symbols

\defcitealias{MartinezVazquez2015}{Paper~I}

\title[]{Probing the early chemical evolution of the Sculptor dSph with purely old stellar tracers}

\author[C. E. Mart{\'i}nez-V{\'a}zquez et al.]{
C. E. Mart{\'i}nez-V{\'a}zquez$^{1,2}$,\thanks{E-mail: clara.marvaz@gmail.com (CEM-V)}
M. Monelli$^{1,2}$,
C. Gallart$^{1,2}$,
G. Bono$^{3,4}$,
E. J. Bernard$^{5}$,
\newauthor{
P. B. Stetson$^{6}$,
I. Ferraro$^{4}$,
A. R. Walker$^{7}$,
M. Dall'Ora$^{8}$,  
G. Fiorentino$^{9}$, and}
\newauthor{
G. Iannicola$^{4}$}
\\
$^{1}$ Instituto de Astrof{\'i}sica de Canarias (IAC), E-38205 La Laguna, Tenerife, Spain\\ 
$^{2}$ Universidad de La Laguna (ULL), Dpto. Astrof{\'i}sica, E-38206 La Laguna, Tenerife, Spain\\ 
$^{3}$ Dipartimento di Fisica, Universit\`a di Roma Tor Vergata, Via della Ricerca Scientifca 1, 00133 Roma, Italy\\
$^{4}$ INAF-Osservatorio Astronomico di Roma, Via Frascati 33, 00040 Monteporzio Catone, Italy\\
$^{5}$ Laboratoire Lagrange, Observatoire de la C\^ote d'Azur, 06304 Nice Cedex 4, France\\
$^{6}$ Herzberg Astronomy and Astrophysics, National Research Council Canada, 5071 West Saanich Road, Victoria, BC V9E 2E7, Canada\\
$^{7}$ Cerro Tololo Inter-American Observatory, National Optical Astronomy Observatory, Casilla 603, La
Serena, Chile\\
$^{8}$ INAF-Osservatorio Astronomico di Capodimonte, Via Moiariello 16, 80131 Napoli, Italy\\
$^{9}$ INAF-Osservatorio Astronomico di Bologna, via Ranzani 1, 40127, Bologna, Italy
}

\date{Accepted 2016 May 6. Received 2016 May 6; in original form 2016 April 14}

\pubyear{2016}

\begin{document}
\label{firstpage}
\pagerange{\pageref{firstpage}--\pageref{lastpage}}
\maketitle

% Abstract of the paper

\begin{abstract}
We present the metallicity distribution of a sample of 471 RR Lyrae (RRL)
stars in the Sculptor dSph, obtained from the $I$-band Period-Luminosity
relation. It is the first time that the early chemical evolution of a dwarf galaxy
is characterized in such a detailed and quantitative way, using photometric data
alone. We find a broad metallicity distribution (FWHM=0.8 dex) that is peaked at
[Fe/H]$\simeq$--1.90 dex, in excellent agreement with literature values obtained
from spectroscopic data. Moreover, we are able to directly trace the metallicity
gradient out to a radius of $\sim$55$\arcmin$.
We find that in the outer regions (r$>\sim$32\arcmin) the slope of the
metallicity gradient from the RRLs (--0.025 dex arcmin$^{-1}$) is comparable to
the literature values based on red giant (RG) stars. However, in the central
part of Sculptor we do not observe the latter gradients.
This suggests that there is a more metal-rich and/or younger population in Sculptor
that does not produce RRLs. This scenario is strengthened by the observation of a
metal-rich peak in the metallicity distribution of RG stars by other authors, which is
not present in the metallicity distribution of the RRLs within the same central area.
\end{abstract}

% Select between one and six entries from the list of approved keywords.
% Don't make up new ones.
\begin{keywords}
stars: variables: RR Lyrae -- galaxies: evolution -- galaxies: individual: Sculptor dSph -- Local Group -- galaxies: stellar content
\end{keywords}

%%%%%%%%%%%%%%%%%%%%%%%%%%%%%%%%%%%%%%%%%%%%%%%%%%%%%%%%%%%%%%%%%%%%%%%%%%%%%%%%

\section{Introduction}\label{sec:introduction} 

RR Lyrae (RRL) stars are low-mass, core helium burning, radially pulsating stars.
They are primary distance indicators, since they obey well defined 
optical/near-infrared Period--Luminosity (PL) relations. They are relatively 
bright and can currently be observed out to $\sim$2 Mpc 
\citep{DaCosta2010,Yang2014}. They are also firm tracers of old ($>$10 Gyr) 
stellar populations, and can be used to constrain the early evolution 
of the host stellar systems \citep{Bernard2008}. 
In particular, their pulsation properties are tightly connected with the 
metallicity distribution of their parent stellar 
populations \citep[see e.g.,][for a review]{Bono2011}.

Several theoretical and empirical investigations have
focused on the use of pulsation properties to constrain the RRL 
metal content. The most popular are Fourier decomposition 
parameters \citep[such as $\phi_{31}$ among 
others, see e.g.,][and references therein]{Nemec2013} 
and the combination of luminosity, period, and amplitude 
\citep[see \S~5 of][for a detailed review]{Jeffery2011}.
Recently, \citet{Marconi2015} presented a new comprehensive theoretical 
framework for RRL stars including a broad range in stellar masses, 
luminosities and chemical compositions. They provided new optical and NIR 
Period-Luminosity-Metallicity (PLZ) and Period-Wesenheit-Metallicity (PWZ) 
relations with various degrees of metallicity dependence. 
In this investigation, we take advantage of the $I$-band PLZ 
relation to evaluate the metallicity of individual Sculptor RRL stars, thus
constraining its early chemical evolution.

Sculptor is a goldmine for understanding 
galaxy evolution, since it is relatively close to the Milky Way 
\citep[$\mu$=19.62 mag,][hereafter Paper~I]{MartinezVazquez2015}, 
and is the crossroads of several theoretical \citep[e.g.,][]{Salaris2013,Romano2013}, 
photometric (e.g., \citealt{DaCosta1984,Hurley-Keller1999,Majewski1999,Monkiewicz1999,Harbeck2001,
Kaluzny1995,deBoer2011,deBoer2012}; \citetalias{MartinezVazquez2015}) 
and spectroscopic investigations 
\citep[e.g.,][]{Tolstoy2004,Clementini2005,Battaglia2008b,Walker2008,Kirby2009,
Starkenburg2013,Skuladottir2015}. Sculptor is a low-mass dwarf galaxy whose
stellar populations display a relatively narrow age range, but still it
is a quite complex system in its structure and chemical enrichment.

The current sample of RRL stars was discussed by Mart{\'i}nez-V{\'a}zquez et 
al. (submitted), based on a photometric 
($BVI$) catalogue covering a time interval of $\sim$24 years and including 536 RRL stars 
(289 RRab, fundamental; 197 RRc, first overtone; 50 RRd, double mode).

\section{Metallicity distribution}\label{sec:metallicity}

Recent theoretical \citep{Marconi2015} and empirical studies (Braga et al. 2016, in 
prep.) indicate that the $I$-band PLZ relation ($I$-PLZ) is a reliable diagnostic 
to derive the metallicity of individual RRLs. By inverting the $I$-PLZ relation, 
we obtain
\begin{equation}\label{eq:invertedplz}
\mathrm{[Fe/H]}=\dfrac{M_I - b\log{P} - a}{c}
\end{equation}
where the coefficients $a$, $b$ and $c$ are listed in Table~6 of
\citet{Marconi2015}. We use here the absolute $M_I$ magnitudes for
a reddening E(B-V)=0.018 mag \citep{Pietrzynski2008} and the RRL distance modulus 
(19.62 mag) derived in \citetalias{MartinezVazquez2015} from the metal-independent 
PW relations. Note that we use $I$-PLZ relations for 
RRab and RRc variables separately, since they have slightly different slopes. 
To illustrate the reliability of the method and provide an empirical estimate of the 
uncertainty on the inferred metallicities, we applied it to the RRLs of the globular 
cluster Reticulum, which have accurate light curves \citep{Kuehn2013}, low reddening 
(E(B-V)=0.016 mag, \citealt{Schlegel1998}), and metallicity 
([Fe/H]=--1.61$\pm$0.15---\citealt{Mackey2004}---on the \citealt{Carretta2009} scale)
close to that of Sculptor.
Applying eq.~\ref{eq:invertedplz} to the corresponding pulsational properties of 22 RRab+RRc
stars in Reticulum (excluding Blazhko and multi-mode RRLs), we obtain a mean metallicity of 
--1.49 dex and $\sigma$=0.25 dex. Since there is no (known) metallicity spread in this cluster, 
the $\sigma$ is representative of the uncertainty of the technique.
In fact, by selecting the best quality light
curves (11) we improve on these results, obtaining a mean metallicity of --1.60 dex and 
$\sigma$=0.07 dex.

\begin{figure}
	\vspace{-0.5cm}
	\includegraphics[scale=0.5]{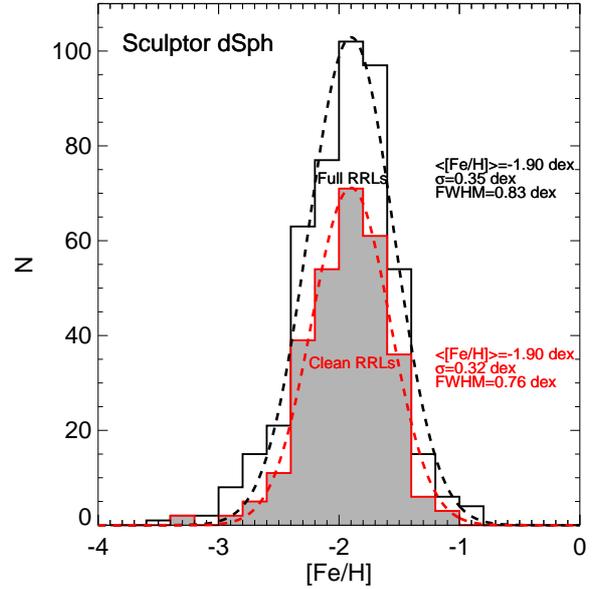}
	\vspace{-0.8cm}
	\caption{Metallicity distribution based on the $I$-PLZ relation for the 
		{\itshape full} (black) and the {\itshape clean} (red and grey filled) 
		samples of RRL stars. The peaks agree well with literature values.}
	\label{fig:metallicity_distribution}
\end{figure}

The black histogram in Fig. \ref{fig:metallicity_distribution} shows the metallicity 
distribution based on the {\itshape full} RRL sample (471 RRab+RRc), while the grey-filled red one 
is for the {\itshape clean} sample (290 RRLs with well sampled and accurate 
light curves, see \citetalias{MartinezVazquez2015} for a detailed discussion 
concerning sample selection).  We fit Gaussian curves
to the metallicity distributions (dashed curves) and 
found that the peak and the $\sigma$ of the two distributions are consistent. 
Therefore, the {\itshape full} RRL sample is adopted in the following analysis.

%%%%%%%%%%%%%%%%%%%%%%%%%%%%%%% FIG 1 %%%%%%%%%%%%%%%%%%%%%%%%%%%%%%%%%%%%%%%%

%%%%%%%%%%%%%%%%%%%%%%%%%%%%%%%%%%%%%%%%%%%%%%%%%%%%%%%%%%%%%%%%%%%%%%%%%%%%

Fig.\ref{fig:metallicity_distribution} shows that the metallicity distribution of
the Sculptor RRL stars peaks at [Fe/H]=--1.90 with $\sigma$=0.35 dex, and a total 
range in metallicity of the order of two dex.
However, the metal-poor and the metal-rich tails of the distribution should be 
treated cautiously since metallicity estimates are based on a photometric index. 
Using the value of $\sigma$ obtained in Reticulum as an
estimate of the dispersion of this method, we find that the intrinsic metallicity spread is
$\sigma\sim0.25\,$dex. On the other hand, we note that recent spectroscopic investigations 
in Sculptor have found both very metal--poor red giant (RG) stars ([Fe/H]$<$--3.0, 
\citealt{Tafelmeyer2010,Starkenburg2013}), and RG stars with [Fe/H]$\sim$--1.0 \citep{Kirby2009}.

%%%%%%%%%%%%%%%%%%%%%%%%%%%%%%% FIG 2 %%%%%%%%%%%%%%%%%%%%%%%%%%%%%%%%%%%%%%%%

\begin{figure*}
	\vspace{-0.8cm} 	
	\includegraphics[scale=0.5]{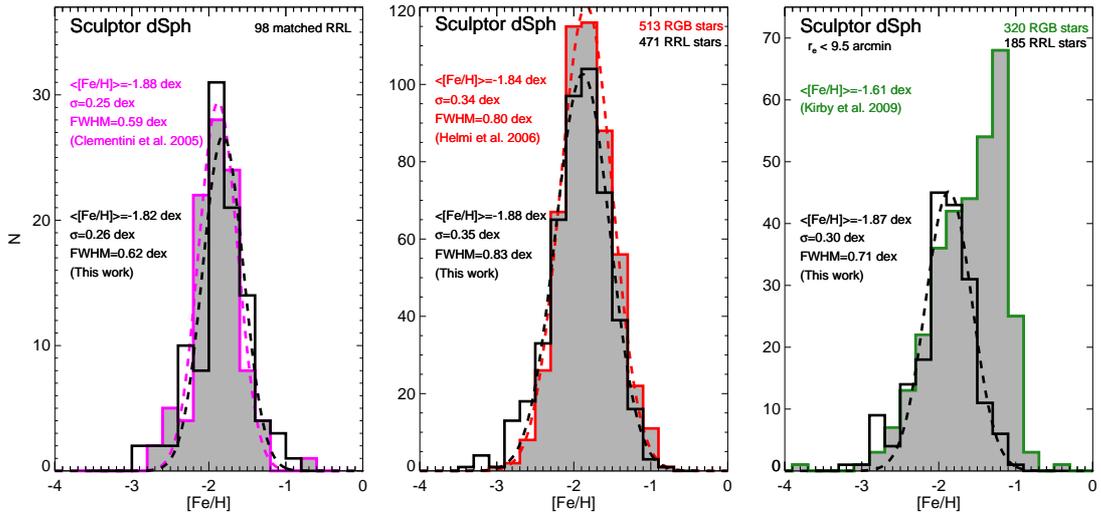}
	\vspace{-0.6cm}
	\caption{Comparison of the derived metallicity distribution with spectroscopic 
		data available in the literature. 
		\emph{Left panel:}
		The 98 RRL stars in common between our sample (black histogram) and that of 
		\citet[][magenta histogram]{Clementini2005}, who obtained the abundances using the $\Delta$S method.
		\emph{Middle panel:} Our full RRL sample (black histogram), vs. the 513 red giant (RG) stars 
		provided by \citet{Helmi2006} using the CaII triplet. 
		\emph{Right panel:} Our RRL stars within 9.5\arcmin, vs. the RG stars provided by \citet{Kirby2009}
		in the same area.
		Note that the above metallicity distributions assume the same solar iron 
		abundance (log $\epsilon$(Fe)=7.52).}
	
	\label{fig:metallicity_comparison}
\end{figure*}

%%%%%%%%%%%%%%%%%%%%%%%%%%%%%%%%%%%%%%%%%%%%%%%%%%%%%%%%%%%%%%%%%%%%%%%%%%%%

Fig. \ref{fig:metallicity_comparison} compares spectroscopic metallicity 
estimates available in the literature. In the left panel, the metallicity distribution 
based on the $I$-PLZ relation (black) is compared to the RRL metallicities 
from \citet[][magenta curves]{Clementini2005}, using a revision of the 
$\Delta$S method (quoted individual errors $\approx \pm$0.15-0.16 dex); we show only 
the 98 RRab+RRc variables in common. For a detailed comparison, both metallicity 
distributions were converted to the homogeneous 
metallicity scale of \citet{Carretta2009}. 
The peaks and $\sigma$'s of the two metallicity distributions are indistinguishable; 
the agreement between the two methods is excellent.

The middle panel shows that the agreement remains good if we 
compare our metallicity distribution with the 
sizeable sample (513 RG stars) of spectroscopic measurements provided by 
\citet[][red histogram]{Helmi2006}. The latter employ low-resolution spectra of
the CaII-triplet as a metallicity diagnostic (quoted error $\approx \pm$0.1 dex). 
Note that the spatial distribution of 
their RG stars and our RRL stars is similar: both include stars out to a radius of 
$\sim$60\arcmin, near the tidal radius of Sculptor. The two metallicity 
distributions are quite symmetric and consistent. 

Finally, the right panel of Fig.~\ref{fig:metallicity_comparison} compares the 
metallicity distribution of our RRL sample with that of the RG stars provided 
by \citet[][green histogram]{Kirby2009} from medium-resolution spectra (quoted error 
$\approx \pm$ 0.15). To limit any bias due to the metallicity gradient 
(see \S~\ref{sec:radialgrad}), the comparison used the RRL and RG
stars located in the region (r$<$9.5\arcmin) homogeneously covered by the 
spectroscopic observations. A glance at the histograms plotted here indicates 
that the RG sample contains a metal-rich component not represented in the 
RRL sample; this component is not obvious in the more spatially extended sample 
of RG from \citet{Helmi2006}, as already noted by \citet{Kirby2009}.
Interestingly, the RRL metallicity distribution does not seem to show significant 
differences between the sample in the innermost region and that covering the 
entire galaxy.

The difference between the metallicity distributions of the RRL and RG stars in 
the centre of Sculptor may be explained by evolutionary effects: 
as the iron abundance increases, the HB morphology becomes 
redder, until the RRL instability strip is no longer populated \citep{Fiorentino2012a}. 
Age has a similar effect, with younger populations also having
a redder HB. This why metal-rich 
globular clusters and younger, outer-halo clusters display a stub of HB red stars 
and very few, if any, RRL stars. This interpretation is supported by the clear difference 
in radial distribution between red and blue HB stars discussed in \citetalias{MartinezVazquez2015}.

%_______________________________________________________________________________
\section{The radial metallicity gradients of the RRL stars in Sculptor}\label{sec:radialgrad}

In  \citetalias{MartinezVazquez2015}, we showed that Sculptor RRLs have
a spread in $V$ magnitude of $\sim$0.35 mag, significantly larger than the 
uncertainties in the mean magnitudes of individual variables ($\sigma$=0.03   
mag), and also significantly larger than the spread expected from evolution 
in a mono-metallic stellar population. We associated the bright (Bt) RRL sample with 
the more metal-poor stellar population, and the faint (Ft) sample with a metal-rich one.

%%%%%%%%%%%%%%%%%%%%%%%%%%%%%%% FIG 3 %%%%%%%%%%%%%%%%%%%%%%%%%%%%%%%%%%%%%%%%

\begin{figure}
  \vspace{-0.5cm}
	\hspace{-0.7cm}
	\includegraphics[scale=0.45]{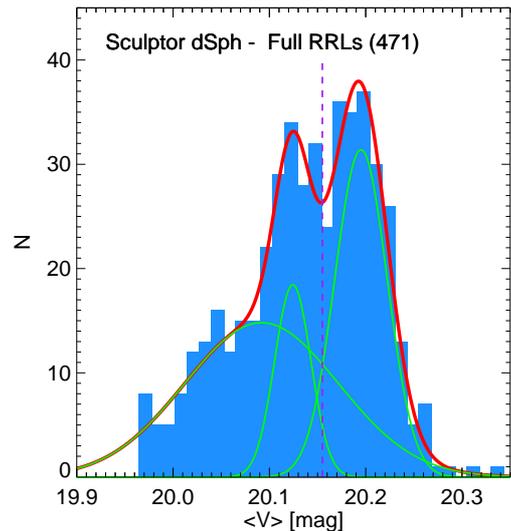}	
	\vspace{-0.5cm}
	\caption{$<V>$-band luminosity function for our RRL sample.
		The green curves display the three individual gaussians adopted to 
		fit the magnitude distribution, while the red curve is the convolution
		of the three gaussians. The purple vertical dashed line marks the luminosity level 
		adopted to separate Bt and Ft RRLs.}
	\label{fig:histo_rrlfull}
\end{figure}

%%%%%%%%%%%%%%%%%%%%%%%%%%%%%%%%%%%%%%%%%%%%%%%%%%%%%%%%%%%%%%%%%%%%%%%%%%%%

Fig.~\ref{fig:histo_rrlfull} shows the $\left<V\right>$-band luminosity function for 
our 471 RRL stars (blue histogram). The green curves 
represent the three Gaussian components adopted to describe the magnitude distribution, while the 
red curve is the sum of the three Gaussians. There are two prominent peaks 
at $\langle V \rangle \sim$20.12 and $\langle V \rangle \sim$20.19. 
We adopt the local minimum at $\left<V\right>$=20.155 (purple vertical dashed line) to split the RRL sample 
into bright (Bt) and faint (Ft) subsamples (\citetalias{MartinezVazquez2015}). 

Taking advantage of the new metallicity determinations we investigate 
the difference in metallicity between the two sub-populations. As expected, the Bt 
sub-sample is, on average, more metal-poor than the Ft one. 
A Gaussian fit to the metallicity distribution of each sample discloses that
the two distributions have similar $\sigma$'s ($\sigma_{\rm Bt}$=0.29 dex, $\sigma_{\rm Ft}$=0.32 dex) 
and partially overlap, but the difference in the mean metallicity 
([Fe/H]=--2.03 [Bt] vs [Fe/H]=--1.74 [Ft]), of the order of 1$\sigma$ (0.3 dex), 
robustly indicates the presence of two sub-populations. 
In \citetalias{MartinezVazquez2015} we showed that the Bt and Ft samples 
follow different spatial distributions, with the Ft RRLs more centrally concentrated 
than the Bt ones, suggesting the presence of a metallicity gradient. \citep{Tolstoy2004} 
reached a similar conclusion from the spatial distribution of red and blue HB stars. 

\begin{figure}
  \vspace{-1cm}
	\hspace{-0.7cm}
	\includegraphics[scale=0.6]{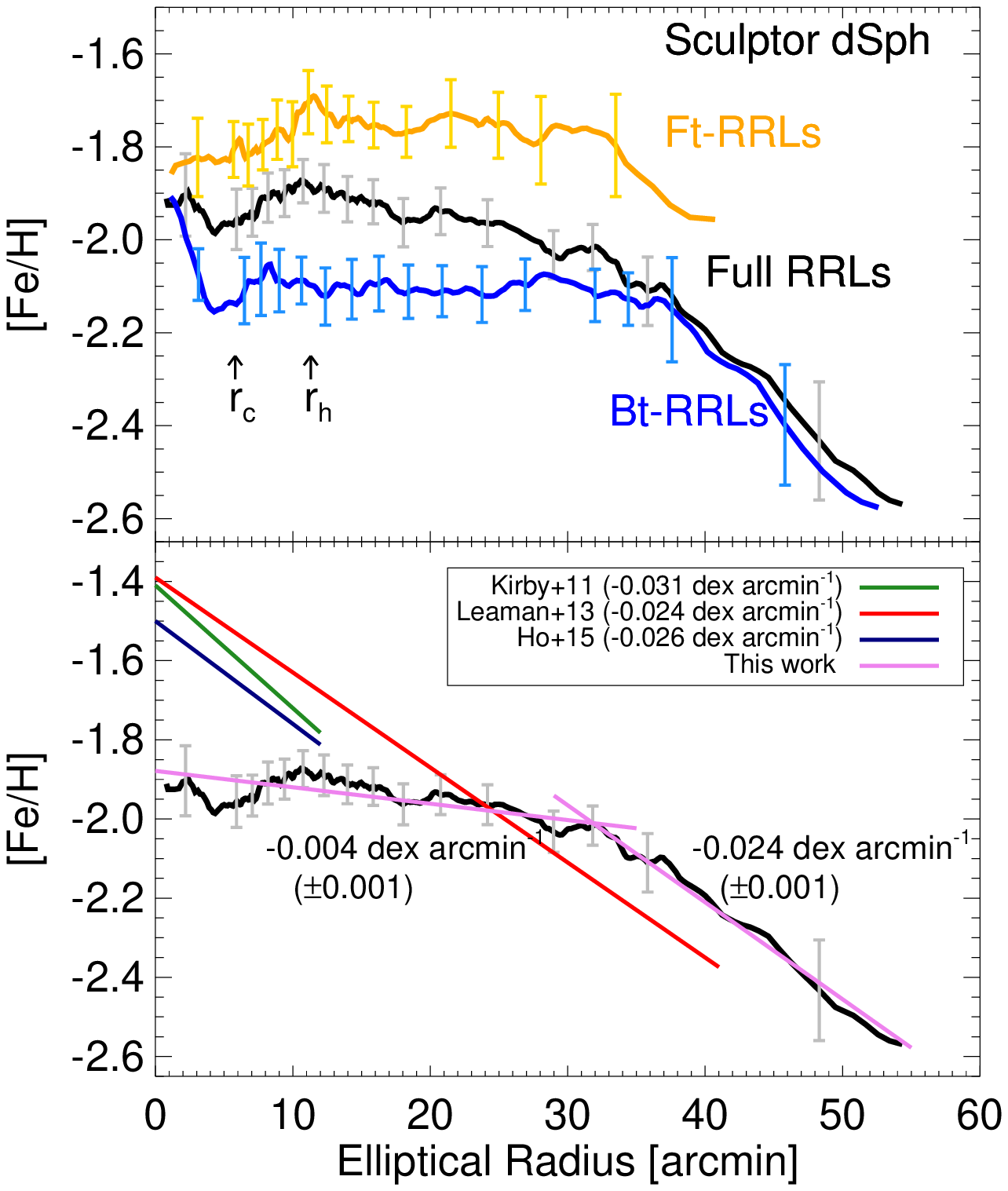}
  \vspace{-0.9cm}
	\caption{\emph{Top}. Metallicity distribution as a function of the elliptical radius for the
		{\itshape full} sample of RRLs (black), Bt-RRLs (blue), and Ft-RRLs (orange).
		The vertical arrows mark the core (r$_c$=5.8\arcmin,\citealt{Mateo1998}) and the half-light
		radius (r$_h$=11.3\arcmin, \citealt{Irwin1995}. The vertical bars display the 
		errors on [Fe/H] based on Monte Carlo simulations. \emph{Bottom}. Comparison with the radial 
		spectroscopic gradients measured by \citet[][green]{Kirby2011}, 
		\citet[][red]{Leaman2013}, and \citet[][dark blue]{Ho2015}. The current and the literature 
		slopes are also labelled.}
	\label{fig:radial_bf}
\end{figure}

Our individual metallicities also allow us
to measure the radial metallicity profile of the oldest 
stellar population in Sculptor. The top panel of Fig. \ref{fig:radial_bf} shows 
the metallicity of the Bt (blue), Ft (orange) and global (black) RRL samples 
as a function of the elliptical radius. The metallicity of the Bt sub-sample 
remains constant, within the errors, over a substantial fraction of the body 
of the galaxy (r$\sim$30\arcmin)---the peak in the centre is not significant 
as shown by the single error bar. At larger radii the profile shows
a steady decrease and approaches a metallicity $\sim$--2.5 dex in the 
outskirts of the galaxy (r$\sim$50\arcmin).
The radial trend of the Ft sub-group is consistent with a constant metallicity 
([Fe/H]$\sim$--1.75), although there is a hint of a lower metallicity within 
the half-light radius (r$_{h}$=11.3\arcmin). The metallicity decrease of the global sample
beyond $\sim$35\arcmin\ is significant, though the error bars are larger due to the
smaller number of RRL stars in the last radial bins. Interestingly, the metallicity 
of the global sample also shows a mild gradient within $\sim$25$\arcmin$, where the 
sub-samples are flat; this is due to the varying ratio of Bt and Ft as a function of radius.

To characterize the change in iron abundance as a function 
of radius more quantitatively, we fit the gradient with two linear relations: 
the violet lines plotted in the bottom panel of Fig.~\ref{fig:radial_bf} show 
the fit for radial distances smaller and larger than r$\sim$32\arcmin.  We find that 
the metallicity profile in the inner region has a mild negative slope 
(-0.004$\pm$0.001 dex/arcmin), while the outermost region displays 
a well defined gradient with a steeper negative slope (-0.025$\pm$0.001 dex/arcmin). 

%%%%%%%%%%%%%%%%%%%%%%%%%%%%%%% FIG 4 %%%%%%%%%%%%%%%%%%%%%%%%%%%%%%%%%%%%%%%%
%%%%%%%%%%%%%%%%%%%%%%%%%%%%%%%%%%%%%%%%%%%%%%%%%%%%%%%%%%%%%%%%%%%%%%%%%%%%

The RRL metallicity gradient of the outermost regions agrees quite well with the 
RG metallicity gradient found by \citet{Leaman2013} from low-resolution spectra 
collected by \citet{Tolstoy2004}. Their sample covers a significant 
fraction of the galaxy and yields a gradient of 
--0.14$\pm$0.02 dex/r$_c$ (--0.024 dex/arcmin assuming r$_c$ equal to 
5.8\arcmin \citep{Mateo1998}---see the red line in the bottom 
panel of Fig.\ref{fig:radial_bf}). However, the quoted gradient is systematically 
steeper than that derived here for radii smaller than 
25\arcmin. In fact, the constant slope of --0.024 dex arcmin$^{-1}$ quoted by \citet{Leaman2013}
implies a central metallicity of --1.4 dex, compared with the --1.9 dex for the average
metallicity of the RRL stars. This evidence further supports the difference between 
the metallicity distribution of purely old 
(RRLs) and possibly mixed-age (RG) stellar tracers discussed in \S\ref{sec:metallicity}.  

The same conclusion applies to the metallicity gradient estimated by \citet{Kirby2011}  
of --0.18$\pm$0.02 dex/r$_c$ (--0.031 dex arcmin$^{-1}$), but the range in radii 
covered by these data is quite limited ($\sim$2 r$_c$, $\sim$12\arcmin). 
Using the same spectra but a different calibration for the CaII triplet,
\citet{Ho2015} found a slightly shallower slope: --0.293$\pm$0.025 dex/r$_h$ 
(--0.026 dex arcmin$^{-1}$). 

We have already mentioned that the current data do not allow us to constrain whether 
the lack of metal-rich RRL (as metal rich as the metal-rich RG from \citealt{Kirby2009}) 
is a consequence of these stars being too metal-rich, too young, or both, since the
effect is the same: the HB is too red to put stars in the instability strip.
However, the metallicity gradient in the RRL indicates that the outermost region in Sculptor 
is dominated by an old stellar population that is more metal-poor than the old stellar
population in the innermost regions. This implies a larger chemical enrichment in the inner
region (by $\sim$0.5 dex) during the period when the stellar population producing the RRL stars
was formed.  The above results also suggest that there is an additional 
stellar population located in the innermost regions that experienced additional enrichment,
causing a drift of the mean metallicity into the metal-intermediate regime ([Fe/H]$\ga$--1.5). 
Under the assumption of continuous metal enrichment, this 
could be a younger population that has no RRL counterpart. This suggests that
star formation and in turn, chemical enrichment lasted longer in the centre of Sculptor 
than in the outermost regions.

%_______________________________________________________________________________
\section{Final remarks}\label{sec:conclusions}

We investigated the metallicity distribution of a sample of 471 RRL
stars in the Sculptor dSph galaxy, based on the $I$-band PLZ relation.
The RRL stars cover a broad range in metallicity ($\sim$2.0 dex),
confirming previous suggestions of
efficient early chemical enrichment in the oldest stellar population in Sculptor. 
We have shown that RRL stars trace a well defined radial metallicity gradient, 
shallow within 32$\arcmin$ but much steeper beyond that.

In particular, we found that in the outer regions (r$\ga$32\arcmin) the slope of the 
metallicity gradient from our RRL sample (--0.025 dex arcmin$^{-1}$) 
is almost the same as the slope derived by \citet{Leaman2013}, who studied 
RG stars out to large galactocentric radius. However, in the central 
part (r$\la$32\arcmin), we do not observe the steep gradients found from 
spectra of RG stars. The gradient in the RRL sample (--0.004 dex arcmin$^{-1}$) is
significantly shallower than the literature values for RGs in the same region 
\citep{Kirby2011,Leaman2013,Ho2015}. This suggests that there is a metal-rich 
and/or younger population in the centre of Sculptor that does not have a counterpart
among the RRLs. This scenario is supported by a metal-rich peak in the 
metallicity distribution of RG stars \citep{Kirby2009} that is not present in the metallicity 
distribution of the RRLs (inside the same area, r$_e <$9.5\arcmin). 

We have presented a powerful new method, from the inverse of the $I$-PLZ relation, 
for studying the internal characteristics of old stellar populations 
in galaxies. The different metallicity profiles of the RRL and RG stars 
in Sculptor reveal a general outside-in formation scenario routinely found in dwarf 
galaxies \citep{Harbeck2001, Stinson2006, Hidalgo2013} but are able to more narrowly delimit 
the successive stellar generations 
in space and time. Beyond $\approx$32\arcmin (5.5 $\times$ $r_c$) from the centre, the RG and RRL  
appear to define a uniform population in terms of age and metallicity. Conversely, 
inside this radius we find solid indications that the RG population is progressively 
more metal-rich and/or younger toward the centre of the galaxy. Under the 
reasonable assumption of a monotonic increase of metallicity with time,  this indicates
a more extended period of star formation toward the centre of Sculptor. Star 
formation must have ended around 10 Gyr ago (the minimum age canonically adopted for RRL stars) 
beyond a radius of $\approx$32\arcmin. Inside this radius, it continued for some time, with younger 
stars becoming progressively more chemically enriched (by up to $\sim$0.5 dex). This scenario agrees
with the spatially resolved SFR(t) obtained by \citet{deBoer2012} from a CMD reaching the oldest main
sequence turnoffs. This shows that RRL stars can provide detailed information on the oldest population 
of a galaxy in the absence of old main-sequence turnoff (MSTO) photometry. In particular, 
by combining RRL and RG metallicities and metallicity profiles, we are able to identify a 
characteristic radius beyond which the stellar population is purely old ($>$10 Gyr) and 
inside which star formation and chemical enrichment proceeded for a longer period of time. 
To quantify the extension of this period to younger ages, the red-HB and red-clump morphology 
can provide some further insight. However, MSTO information is still the best bet for accurate ages.

\vspace{-0.7cm}
\section*{Acknowledgments}

The authors thank the anonymous referee for the useful comments.
This research has been supported by the Spanish Ministry of Economy and 
Competitiveness (MINECO) under the grant (project reference AYA2014-56795-P).
EJB acknowledges support from the CNES postdoctoral fellowship program.
GF has been supported by the Futuro in Ricerca 2013 (grant RBFR13J716).

\vspace{-0.5cm}
%%%%%%%%%%%%%%%%%%%%%%%%%%%%%%%%%%%%%%%%%%%%%%%%%%

%%%%%%%%%%%%%%%%%%%% REFERENCES %%%%%%%%%%%%%%%%%%

% The best way to enter references is to use BibTeX:

\bibliographystyle{mnras}

%\bibliography{example} % if your bibtex file is called example.bib
%\bibliography{cmartinez_bibtex}
 \newcommand{\noop}[1]{}

%\appendix
%
%\newpage

%\subsection{Subsection title}

%%%%%%%%%%%%%%%%%%%%%%%%%%%%%%%%%%%%%%%%%%%%%%%%%%

% Don't change these lines
\bsp	% typesetting comment
\label{lastpage}
\end{document}